\documentclass{article}

\usepackage{amsmath}
\usepackage{amstext}
\usepackage{amssymb}
\usepackage{revsymb}
\usepackage{latexsym}

\begin{document}
\title{The Heisenberg Uncertainty Principle and the\\
Nyquist-Shannon Sampling Theorem}
\author{Pierre A. Millette\thanks{Currently with EMS Technologies, Ottawa,
CANADA}}

\maketitle

\begin{abstract}
The derivation of the Heisenberg Uncertainty Principle (HUP) from the
Uncertainty Theorem of Fourier Transform theory demonstrates that
the HUP arises from the dependency of momentum on wave number that
exists at the quantum level.  It also establishes that the HUP is
purely a relationship between the effective widths of Fourier transform
pairs of variables (i.e. conjugate variables).  We note that
the HUP is not a quantum mechanical measurement principle per se.
We introduce the Quantum Mechanical equivalent of the Nyquist-Shannon
Sampling Theorem of Fourier Transform theory, and show that
it is a better principle to describe the measurement limitations
of Quantum Mechanics.  We show that Brillouin zones in Solid State
physics are a manifestation of the Nyquist-Shannon Sampling Theorem
at the quantum level.  By comparison with other fields where Fourier
Transform theory is used, we propose that we need to discern between
measurement limitations and inherent limitations when interpreting
the impact of the HUP on the nature of the quantum level.  We
further propose that while measurement limitations result in our
perception of indeterminism at the quantum level, there is no
evidence that there are any inherent limitations at the quantum
level, based on the Nyquist-Shannon Sampling Theorem.
\end{abstract}

\section{Introduction} \label{Intro}

The Heisenberg Uncertainty Principle is a cornerstone of quantum
mechanics.  As noted by Hughes \cite{hughes}, the
interpretation of the Principle varies

\begin{itemize}

\item from expressing a limitation on measurement as originally
derived by Heisenberg \cite{heisenberg} (Heisenberg's microscope),

\item to being the variance of a measurement carried out on an
ensemble of particles \cite{popper} \cite{margenau},

\item to being inherent to a microsystem \cite{davies}, meaning
essentially that there is an indeterminism to the natural world
which is a basic characteristic of the quantum level.

\end{itemize}
Greenstein retains only the first and last alternatives
\cite{greenstein1}.

However, the Heisenberg Uncertainty Principle can be derived from
considerations which clearly demonstate that these interpretations
of the principle are not required by its mathematical formulation.
This derivation, based on the application of Fourier methods, is
given in various mathematical and engineering books, for example
\cite{cartwright}.

\section{Consistent Derivation of the \\Heisenberg Uncertainty Principle} \label{HUP}

In the Fourier transform literature, the Heisenberg Uncertainty
Principle is derived from a general theorem of Fourier theory
called the Uncertainty Theorem \cite{cartwright}. This
theorem states that the effective width of a function times the
effective width of its transform cannot be less than a minimum
value given by:
\begin{equation} \label{WW}
W(f) \cdot W(\tilde{f}) \geq 1/2
\end{equation}
where $f$ is the function of interest and $\tilde{f}$ is its
Fourier transform.  $W(f)$ is the effective width of function $f$,
defined by
\begin{equation} \label{W}
|W(f)|^{2} = \frac{\int_{-\infty}^{\infty} |f(u)|^2 [u-M(f)]^2
du}{\int_{-\infty}^{\infty} |f(u)|^2 du}
\end{equation}
 and $M(f)$ is the mean ordinate defined by
\begin{equation} \label{M}
M(f) = \frac{\int_{-\infty}^{\infty} |f(u)|^2 u
du}{\int_{-\infty}^{\infty} |f(u)|^2 du}
\end{equation}

There are several points that must be noted with respect to this
derivation:

\begin{itemize}

\item Equation (\ref{WW}) applies to a Fourier transform pair of
variables. Taking the simple case of time $t$ and frequency $\nu$
to illustrate the point: If we consider the function $f$ to be the
function that describes a time function $t$, then the width of the
function, $W(f)$, can be denoted as $W(f) = \Delta t$. The Fourier
transform of function $t$ is the frequency function $\nu$ and the
width of this function can be denoted as $W(\tilde{t}) = W(\nu)
= \Delta \nu$. Substituting in Equation (\ref{WW}), the
Uncertainty Theorem then yields
\begin{equation} \label{dtdf}
\Delta t \cdot \Delta \nu \geq 1/2
\end{equation}

\item However, if one wishes to use the circular frequency $\omega
= 2 \pi \nu$ instead, Equation (\ref{dtdf}) becomes
\begin{equation} \label{dtdo}
\Delta t \cdot \Delta \omega \geq \pi
\end{equation}

\end{itemize}
It is thus necessary to take special care to clearly identify the
Fourier transform variable used as it impacts the R.H.S. of the
resulting Uncertainty relation (see for example
\cite{weisstein-FT} and \cite{griffiths}).

Equations (\ref{dtdf}) and (\ref{dtdo}) above correspond to the
following definitions of the Fourier transform
\cite{weisstein-FT}:

\begin{itemize}

\item Equation (\ref{dtdf}):
\begin{equation} \label{f} f(t) = \int_{-\infty}^{\infty}
\tilde{f}(\nu) \exp(2\pi i\nu t) d\nu
\end{equation}
\begin{equation} \label{ftilde}
\tilde{f}(\nu) = \int_{-\infty}^{\infty} f(t) \exp(-2\pi i\nu t)
dt
\end{equation}

\item Equation (\ref{dtdo}):
\begin{equation} \label{g} g(t) = \frac{1}{2\pi}
\int_{-\infty}^{\infty} \tilde{g}(\omega) \exp(i\omega t) d\omega
\end{equation}
\begin{equation} \label{gtilde}
\tilde{g}(\omega) = \int_{-\infty}^{\infty} g(t) \exp(-i\omega t)
dt
\end{equation}

\end{itemize}
respectively.  Sometimes the factor $\frac{1}{2\pi}$ is
distributed between the two integrals (the Fourier and the Inverse
Fourier Transform Integrals) as $\frac{1}{\sqrt{2\pi}}$.  In
Physics, Equations (\ref{g}) and (\ref{gtilde}) are preferred, as
this eliminates the cumbersome factor of $2\pi$ in the exponential
(see for example \cite{ziman1}), but care must then be taken to ensure the
resulting factor of $\frac{1}{2\pi}$ in Equation (\ref{g}) is
propagated forward in derivations using that definition.

Using the relation $E = h \nu$, where $h$ is Planck's constant, in
Equation (\ref{dtdf}) above, or the relation $E = \hbar \omega$,
where $\hbar = h / 2\pi$, in Equation (\ref{dtdo}) above, one
obtains the same statement of the Heisenberg Uncertainty Principle
namely
\begin{equation} \label{dtdE}
\Delta E \cdot \Delta t \geq h/2
\end{equation}
in both cases.

Similarly for the position $x$, if we consider the function $f$ to
be the function that describes the position $x$ of a particle,
then the width of the function, $W(f)$, can be denoted as $W(f) =
\Delta x$. The Fourier transform of function $x$ is the function
$\tilde{x} = \lambda^{-1}$ and the width of this function can be
denoted as $W(\tilde{x}) = W(\lambda^{-1}) = \Delta
(\lambda^{-1})$ which we write as $\Delta \lambda^{-1}$ for
brevity. You will note that we have not used the wavenumber
function $k$, as this is usually defined as $k = 2 \pi / \lambda$
(see for example \cite{weisstein-k} and references). Substituting
in Equation (\ref{WW}), we obtain the relation
\begin{equation} \label{dxdl}
\Delta x \cdot \Delta \lambda^{-1} \geq 1/2
\end{equation}
In terms of the wavenumber $k$, Equation (\ref{dxdl}) becomes
\begin{equation} \label{dxdk}
\Delta x \cdot \Delta k \geq \pi
\end{equation}

Given that the momentum of a quantum particle is given by $p = h /
\lambda$ or by $p = \hbar k$, both Equations (\ref{dxdl}) and
(\ref{dxdk}) can be expressed as
\begin{equation} \label{dxdp}
\Delta x \cdot \Delta p \geq h/2
\end{equation}
Equations (\ref{dtdE}) and (\ref{dxdp}) are both different statements of the
Heisenberg Uncertainty Principle.

The R.H.S. of these equations is different from the usual
statement of the Heisenberg Uncertainty Principle where the value
$\hbar/2$ is used instead of the value $h/2$ obtained in this
analysis. The application of Equation (\ref{dtdf}) to circular
variables (i.e. using $\omega$ in Equation (\ref{dtdf}) instead
of Equation (\ref{dtdo})) would result in the (incorrect) expression
\begin{equation} \label{dtdowrong}
\Delta t \cdot \Delta \omega \geq 1/2
\end{equation}
and the more commonly encountered expression
\begin{equation} \label{dtdEwrong}
\Delta E \cdot \Delta t \geq \hbar/2
\end{equation}

However, Heisenberg's original derivation \cite{heisenberg} had the R.H.S.
of Equation (\ref{dxdp}) approximately equal to $h$, and Greenstein's
rederivation \cite{greenstein2} of Heisenberg's principle results
in the value $h/2$.  Kennard's formal derivation \cite{kennard} using standard
deviations established the value of $\hbar/2$ used today. This would
thus seem to be the reason for the use of the value $\hbar/2$ in
the formulation of the Heisenberg Uncertainty Principle.

Recently, Sch\"{u}rmann et al \cite{schurmann} have shown that in the case of
a single slit diffraction experiment, the standard deviation
of the momentum typically does not exist. They derive the conditions
under which the standard deviation of the momentum is finite, and show that
the R.H.S. of the resulting inequality satisfies Equation (\ref{dxdp}).
It thus seems that Equation (\ref{dxdp}) is the more general formulation
of the Heisenberg Uncertainty Principle, while the expression with the value
$\hbar/2$ derived using standard deviations is a more specific case.

Whether one uses $\hbar/2$ or
$h/2$ has little impact on the Heisenberg Uncertainty Principle as
the R.H.S. is used to provide an order of magnitude estimate of
the effect considered.  However, the difference becomes evident
when we apply our results to the Brillouin zone formulation of
Solid State Physics (as will be seen in Section (\ref{Interp}))
since this now impacts calculations resulting from models that can
be compared with experimental values.

\section{Interpretation of the Heisenberg Uncertainty Principle} \label{IHUP}

This derivation demonstrates that the Heisenberg Uncertainty
Principle arises because $x$ and $p$ form a Fourier transform pair
of variables. It is a characteristic of Quantum Mechanics that
conjugate variables are Fourier transform pairs of variables. Thus
the Heisenberg Uncertainty Principle arises because the momentum
$p$ of a quantum particle is proportional to the de Broglie wave
number $k$ of the particle. If momentum was not proportional to
wave number, the Heisenberg Uncertainty Principle would not exist
for those variables.

This argument elucidates why the Heisenberg Uncertainty Principle
exists. Can it shed light on the meaning of the Heisenberg
Uncertainty Principle in relation to the basic nature of the
quantum level? First, we note that the Uncertainty Principle,
according to Fourier transform theory, relates the effective width
of Fourier transform pairs of functions or variables. It is not a
measurement theorem per se. It does not describe what happens when
Fourier tranform variables are measured, only that their effective
widths must satisfy the Uncertainty Principle.

Indeed, as pointed out by Omn\`{e}s \cite{omnes}, "it is
quite legitimate to write down an eigenstate of energy at a
well-defined time".  Omn\`{e}s ascribes this seeming violation of
the Heisenberg Uncertainty Principle to the fact that time is not
an observable obtained from an operator like momentum, but rather
a parameter. Greenstein \cite{greenstein3} makes the same
argument. However, time $t$ multiplied by the speed of light $c$
is a component of the 4-vector $x^\mu$ and energy $E$ divided by $c$
is a component of the energy-momentum 4-vector $p^\mu$.  The time
component of these 4-vectors should not be treated differently than
the position component.  The operator versus parameter argument is
weak.

What Omn\`{e}s' example shows is that the impact of the effective
widths $\Delta t$ and $\Delta E$ of the Heisenberg Uncertainty Principle
depends on the observation of the time function $t$ and of the energy function
$E$ that is performed.  A time interval $\Delta t$ can be associated with the
time function $t$ during which is measured the energy eigenstate function
$E$ which itself has a certain width $\Delta E$, with both $\Delta$'s satisfying
Equation (\ref{dtdE}).  This example demonstrates that the Heisenberg Uncertainty
Principle is not a measurement theorem as often used.  Rather, it is a
relationship between the effective widths of Fourier transform pairs of
variables that can have an impact on the observation of those variables.

A more stringent scenario for the impact of the energy-time Heisenberg
Uncertainty Principle is one where the time and energy functions are
small quantities.  For example, we consider the impact of $\Delta t$ on
the observation of $\tau_n$, the lifetime of an atom in energy eigenstate
$n$, and the impact of $\Delta E$ on the transition energy $E_{mn}$, for
a transition between states $n$ and $m$ during spectral line emission.
The conditions to be able to observe $\tau_n$ and $E_{mn}$ are:
\begin{equation} \label{tE0a}
\tau_n \geq \Delta t
\end{equation}
\begin{equation} \label{tE0b}
E_{mn} \geq \Delta E
\end{equation}
Using Equation (\ref{dtdE}) in Equation (\ref{tE0a}) above,
\begin{equation} \label{tE1}
\tau_n \geq \Delta t \geq h/(2\Delta E)
\end{equation}
Hence
\begin{equation} \label{tE2}
\Delta E \geq \frac{h}{2} \frac{1}{\tau_n}
\end{equation}
As state $n$ increases, the lifetime $\tau_n$ decreases.  Equation (\ref{tE2})
is thus more constrained in the limit of large $n$.  Using
the following hydrogenic asymptotic expression for $\tau_n$ from
Millette et al \cite{millette}
\begin{equation} \label{tE3}
\tau_n \thicksim \frac{n^5}{\ln(n)}
\end{equation}
into Equation (\ref{tE2}), Equation (\ref{tE0b}) becomes
\begin{equation} \label{tE4}
E_{mn} \geq \Delta E \gtrsim \frac{h}{2} \text{ } k \frac{\ln(n)}{n^5}
\end{equation}
where $1/k$ is the constant of proportionality of Equation (\ref{tE3}) given by
\begin{equation} \label{tEk}
k = \frac{2^6}{3} \sqrt \frac{\pi}{3} Z^2 \alpha^3 c R_H
\end{equation}
where $Z$ is the nuclear charge of the hydrogenic ion, $\alpha$ is the fine-structure constant,
and $R_H$ is the hydrogen Rydberg constant.
Eliminating the middle term, Equation (\ref{tE4}) becomes
\begin{equation} \label{tE4b}
E_{mn} \gtrsim \frac{h}{2} \text{ } k \frac{\ln(n)}{n^5}
\end{equation}
Applying L'H\^{o}pital's rule, the R.H.S. of the above equation is of order
\begin{equation} \label{tE5}
\text{R.H.S. }\thicksim \text{ } O(\frac{1}{n^5}) \text{  as n} \rightarrow \infty
\end{equation}
while the L.H.S. is of order \cite{bethe}
\begin{equation} \label{tE6}
\text{L.H.S. }\thicksim \text{ } O(\frac{1}{n^2}) \text{  as n} \rightarrow \infty
\end{equation}
Given that Equation (\ref{tE5}) tends to zero faster than Equation (\ref{tE6}),
Equation (\ref{tE4b}) is satisfied.
Both $\tau_n$, the lifetime of the atom in energy eigenstate $n$, and the transition
energy $E_{mn}$ for the transition between states $n$ and $m$ satisfy the conditions
for observation of the spectral line emission.  Thus for the time interval $\Delta t$,
given by Equation (\ref{tE0a}), associated with the time function $\tau_n$ for the
transition energy function $E_{mn}$ which itself has a certain width $\Delta E$,
given by Equation (\ref{tE0b}), both $\Delta$'s satisfy
Equation (\ref{dtdE}) as expected, given the observation of spectral line emission.

\section{Quantum Measurements and the \\Nyquist-Shannon Sampling
Theorem} \label{NSST}

At the quantum level, one must interact to some degree with a
quantum system to perform a measurement.  When describing the
action of measurements of Fourier transform variables, one can
consider two limiting measurement cases:  1) truncation of the
variable time series as a result of a fully interacting
measurement or 2) sampling of the variable time series at intervals
which we consider to be regular in this analysis, in the case
of minimally interacting measurements. As we
will see, the action of sampling allows for measurements that
otherwise would not be possible in the case of a single minimal
interaction.

It should be noted that the intermediate case of a
partial measurement interaction resulting for example in a
transfer of energy or momentum to a particle can be considered as
the truncation of the original time series and the initiation of a
new time series after the interaction. The advantage of
decomposing measurement actions in this fashion is that their
impact on Fourier transform variables can be described by the
Nyquist-Shannon Sampling Theorem of Fourier transform theory. This
theorem is a measurement theorem for Fourier transform variables
based on sampling and truncation operations.

The Nyquist-Shannon Sampling Theorem is fundamental to the field
of information theory, and is well known in digital signal
processing and remote sensing \cite{oppenheim}. In its most basic
form, the theorem states that the rate of sampling of a signal (or
variable) $f_{s}$ must be greater than or equal to the Nyquist sampling
rate $f_{S}$ to avoid loss of information in the sampled signal,
where the Nyquist sampling rate is equal to twice that of the
highest frequency component, $f_{max}$, present in the signal:
\begin{equation} \label{fNyquist}
f_{s} \geq f_{S} = 2 f_{max}.
\end{equation}
If the sampling rate is less than that of Equation
(\ref{fNyquist}), aliasing occurs, which results in a loss of
information.

In general, natural signals are not infinite in duration and,
during measurement, sampling is also accompanied by truncation of
the signal. There is thus loss of information during a typical
measurement process. The Nyquist-Shannon Sampling theorem
elucidates the relationship between the process of sampling and
truncating a variable and the effect this action has on its
Fourier transform \cite{brigham}. In effect, it explains
what happens to the information content of a variable when its
conjugate is measured.

Sampling a variable $x$ at a rate $\delta x$ will result in the
measurement of its conjugate variable $\tilde{x}$ being limited to
its maximum Nyquist range value $\tilde{x}_N$ as given by the
Nyquist-Shannon Sampling theorem:
\begin{equation} \label{b}
\tilde{x} \leq \tilde{x}_N
\end{equation}
where
\begin{equation} \label{bN}
\tilde{x}_N = 1 / (2 \delta x).
\end{equation}
Combining these two equations, we get the relation
\begin{equation} \label{bfull}
\tilde{x} \cdot \delta x \leq 1/2, \text{ for } \tilde{x} \leq \tilde{x}_N
\end{equation}
Conversely, truncating a variable $x$ at a maximum value $x_N$ ($x
\leq x_N$) will result in its conjugate variable $\tilde{x}$ being
sampled at a rate $\delta \tilde{x}$ given by the Nyquist-Shannon
Sampling theorem $\delta \tilde{x} = 1 / (2 x_N)$ resulting in the
relation
\begin{equation} \label{db}
\delta \tilde{x} \cdot x \leq 1/2, \text{ for } x \leq x_N.
\end{equation}

The impact of the Nyquist-Shannon Sampling theorem is now
considered for a particle's position $x$ and momentum $p$.
Applying the theorem to the case where a particle's trajectory is
truncated to $x_N$, we can write from Equation (\ref{db}), for $x
\leq x_N$,
\begin{equation} \label{xLdl}
x \cdot \delta \lambda^{-1} \leq 1/2, \text{ for } x \leq x_N
\end{equation}
or
\begin{equation} \label{xLdk}
x \cdot \delta k \leq \pi, \text{ for } x \leq x_N
\end{equation}
which becomes
\begin{equation} \label{xLdp}
x \cdot \delta p \leq h/2, \text{ for } x \leq x_N
\end{equation}
where $\delta p$ is the $p$-domain sampling rate and the $x$
values can be measured up to $x_N$ (corresponding to the equality
in the equations above).

Conversely, applying the theorem to the case where a particle's
trajectory is sampled at a rate $\delta x$, one can also write
from Equation (\ref{bfull}), for $\tilde{x} \leq \tilde{x}_N$,
where $\tilde{x}$ stands for either of $\lambda^{-1}$, $k$, or $p$,
\begin{equation} \label{dxlL}
\delta x \cdot \lambda^{-1} \leq 1/2, \text{ for } \lambda^{-1} \leq \lambda^{-1}_N
\end{equation}
or
\begin{equation} \label{dxkL}
\delta x \cdot k \leq \pi, \text{ for } k \leq k_N
\end{equation}
which becomes
\begin{equation} \label{dxpL}
\delta x \cdot p \leq h/2, \text{ for } p \leq p_N
\end{equation}
where $\delta x$ is the $x$-domain sampling rate and $k_N$ is the
wave number range that can be measured. For the case where the
equality holds, we have $k_N = \pi / \delta x$ where $k_N$ is the
Nyquist wave number, the maximum wave number that can be measured
with a $\delta x$ sampling interval.

Sampling in one domain leads to truncation in the other. Sampling
($\delta x$) and truncation ($x_N$) in one domain leads to
truncation ($k_N$) and sampling ($\delta k$) respectively in the
other. As $x$ and $k$ form a Fourier transform pair in quantum
mechanics, the Nyquist-Shannon Sampling theorem must also apply to
this pair of conjugate variables.  Similar relations can be
derived for the $E$ and $\nu$ pair of conjugate variables.

\section{Implications of the Nyquist-Shannon Sampling
Theorem at the Quantum Level} \label{Interp}

Equations (\ref{xLdk}) and (\ref{dxkL}) lead to the following
measurement behaviors at the quantum level:

\begin{itemize}

\item Lower-bound limit:  If the position of a particle is
measured over an interval $x_N$, its wave number cannot be
resolved with a resolution better than sampling rate $\delta k$ as
given by Equation (\ref{xLdk}) with $x = x_N$. If the momentum of
a particle is measured over an interval $k_N$, its position cannot
be resolved with a resolution better than sampling rate $\delta x$
as given by Equation (\ref{dxkL}) with $k = k_N$.

\item Upper-bound limit:  If the position of a particle is sampled
at a rate $\delta x$, wave numbers up to $k_N$ can be
resolved, while wave numbers larger than $k_N$ cannot be
resolved as given by Equation (\ref{dxkL}). If the momentum of a
particle is sampled at a rate $\delta k$, lengths up to
$x_N$ can be resolved, while lengths longer than
$x_N$ cannot be resolved as given by Equation (\ref{xLdk}).

\end{itemize}

The lower-bound limit is similar to how the Heisenberg Uncertainty Principle
is usually expressed when it is used as a measurement principle, although
it is not strictly equivalent.
The Nyquist-Shannon Sampling Theorem provides the proper
formulation and limitations of this type of measurement.

The upper-bound limit suggests a different type of quantum measurement:
regular sampling of a particle's position or momentum.
In this case, one can obtain as accurate a
measurement of the Fourier transform variable as desired, up to
the Nyquist-Shannon Sampling limit of $h/2$ (i.e. in the interval $[0 , h/2]$).

An example of this phenomenon occurs in Solid State physics where the
translational symmetry of atoms in a solid resulting from the
regular lattice spacing, is equivalent to an effective sampling of
the atoms of the solid and gives rise to the Brillouin zone for
which the valid values of $k$ are governed by Equation
(\ref{dxkL}).  Setting $\delta x = a$, the lattice spacing, and
extending by symmetry the $k$ values to include the symmetric negative values,
one obtains \cite{harrison}, \cite{chaikin},
\cite{ziman2}:
\begin{equation} \label{Bzone}
-\pi / a \leq k \leq \pi / a
\end{equation}
or alternatively
\begin{equation} \label{Bzone}
k \leq | \pi / a |.
\end{equation}
This is called the reduced zone scheme and $\pi/a$ is called the
Brillouin zone boundary \cite{kittel}.

In essence, this is a theory of measurement for variables that are
Fourier transform pairs. The resolution of our measurements is
governed by limitations that arise from the Nyquist-Shannon
Sampling theorem. Equations (\ref{xLdk}) and (\ref{dxkL}) are
recognized as measurement relationships for quantum mechanical
conjugate variables. Currently, Quantum Mechanics only considers
the Uncertainty Theorem but not the Sampling Theorem. The two
theorems are applicable to Quantum Mechanics and have different
interpretations: the Uncertainty Theorem defines a relationship
between the widths of conjugate variables, while the Sampling
Theorem establishes sampling and truncation measurement
relationships for conjugate variables.

The value $\delta x$ is a sampled measurement and as a result can
resolve values of $p$ up to its Nyquist value $p_N$ given by the
Nyquist-Shannon Sampling theorem, Equation (\ref{dxpL}).  This is
a surprising result as the momentum can be resolved up to its
Nyquist value, in apparent contradiction to the Heisenberg
Uncertainty Principle.  Yet this result is known to be correct as
demonstrated by the Brillouin zones formulation of Solid State
Physics. Physically this result can be understood from the
sampling measurement operation which builds up the momentum
information during the sampling process, up to the Nyquist limit
$p_N$. It must be remembered that the Nyquist limit depends on the
sampling rate $\delta x$ as per the Nyquist-Shannon Sampling
theorem, Equation (\ref{dxpL}).  The Nyquist value must also
satisfy Equation (\ref{fNyquist}) to avoid loss of information in
the sampling process, due to aliasing.

This improved understanding of the Heisenberg Uncertainty
Principle and its sampling counterpart allows us to clarify its
interpretation.  This is based on our understanding of the behavior
of the Uncertainty Theorem and the Nyquist-Shannon Sampling Theorem
in other applications such as, for example, Digital Signal
Processing.

\section{Measurement Limitations and \\Inherent Limitations} \label{Limitations}

It is important to differentiate between the measurement
limitations that arise from the properties of Fourier transform
pairs previously considered, and any inherent limitations that may
or may not exist for those same variables independently of the
measurement process. Quantum theory currently assumes that the
inherent limitations are the same as the measurement limitations.
This assumption needs to be re-examined based on the improved
understanding obtained from the effect of the Uncertainty and
Sampling Theorems in other applications.

The properties of Fourier transform pairs considered in the
previous sections do not mean that the underlying quantities we
are measuring are inherently limited by our measurement
limitations. On the contrary, we know from experience in other
applications that our measurement limitations do not represent an
inherent limitation on the measured quantities in Fourier
Transform theory: for example, in Digital Signal Processing,
a signal is continuous even though our
measurement of the signal results in discrete and aliased values
of limited resolution subject to the Nyquist-Shannon Sampling
Theorem (analog and digital representation of the signal).  The
effective width of the signal and its transform are related by the
Uncertainty theorem. Even though the time and frequency evolution
of a signal that we measure is limited by our measurement
limitations, the time domain and frequency domain signals are both
continuous, independently of how we measure them.

The measurement limitations apply equally to the macroscopic level
and to the quantum level as they are derived from the properties
of Fourier transform pairs of variables which are the same at all
scales. However, at the quantum level, contrary to our macroscopic
environment, we cannot perceive the underlying quantities other
than by instrumented measurements. Hence during a measurement
process, the quantum level is limited by our measurement
limitations. However, assuming that these measurement limitations
represent inherent limitations and form a basic characteristic of
the quantum level is an assumption that is not justified based on
the preceding considerations. Indeed, the Nyquist-Shannon Sampling
Theorem of Fourier Transform theory shows that the range of values
of variables below the Heisenberg Uncertainty Principle value of $h/2$ is
accessible under sampling measurement conditions, as demonstrated by
the Brillouin zones formulation of Solid State physics.

\section{Overlap of the Heisenberg Uncertainty \\Principle and the
Nyquist-Shannon \\Sampling Theorem} \label{Overlap}

Brillouin zone analysis in Solid State physics demonstrates that
one can arbitrarily measure $k$ from $0$ up to its Nyquist limit,
as long as the variable $x$ is sampled at a constant rate (rather than performing a
single $x$ measurement). The Nyquist-Shannon Sampling Theorem can
thus be considered to cover the range that the Heisenberg Uncertainty
Principle excludes.

However, one should recognize that the coverage results from two disparate
theorems, and one should be careful not to try to tie the two Theorems at
their value of overlap $\pi$.  The reason is that one expression involves
the widths of conjugate variables as determined by Equations (\ref{WW})
to (\ref{M}), while the other involves sampling a variable and
truncating its conjugate, or vice versa as determined by Equations (\ref{xLdk}) and
(\ref{dxkL}).  The equations are not continuous at the point of overlap $\pi$.
Indeed, any relation obtained would apply only at the overlap $\pi$
and would have no applicability or physical validity on either side of the overlap.

\section{Conclusion} \label{Conclusion}

In this paper, we have shown that a consistent application of
Fourier Transform theory to the derivation of the Heisenberg
Uncertainty Principle requires that the R.H.S. of the Heisenberg
inequality be $h/2$, not $\hbar/2$.  This is confirmed when
extending the analysis to the Brillouin zones formulation of
Solid State Physics.

We have noted that the Heisenberg Uncertainty Principle, obtained from the
Uncertainty Theorem of Fourier Transform theory, arises because of
the dependency of momentum on wave number that exists at the
quantum level.  Quantum mechanical conjugate variables are Fourier
Transform pairs of variables.

We have shown from Fourier Transform theory that the
Nyquist-Shannon Sampling Theorem affects the nature of
measurements of quantum mechanical conjugate variables. We have
shown that Brillouin zones in Solid State physics are a
manifestation of the Nyquist-Shannon Sampling Theorem at the
quantum level.

We have noted that both the Sampling Theorem and the Uncertainty
Theorem are required to fully describe quantum mechanical
conjugate variables. The Nyquist-Shannon Sampling Theorem complements
the Heisenberg Uncertainty Principle.  The overlap of
these Theorems at the $h/2$ equality value is a mathematical artifact
and has no physical significance.

We have noted that the Uncertainty Theorem and the Nyquist-Shannon
Sampling Theorem apply to Fourier Transform pairs of variables
independently of the level at which the theorems are applied
(macroscopic or microscopic). Conjugate variable measurement
limitations due to these Theorems affect how we perceive quantum
level events as these can only be perceived by instrumented
measurements at that level. However, based on our analysis,
quantum measurement limitations affect our perception of the
quantum environment only, and are not inherent limitations of the
quantum level, as demonstrated by
the Brillouin zones formulation of Solid State physics.

The application of the Nyquist-Shannon Sampling Theorem to the
quantum level offers the possibility of investigating new
experimental conditions over and above the Brillouin zone example
from Solid State physics considered in this paper, allowing a unique
vista into a range of variable values previously considered
unreachable due to the Heisenberg Uncertainty Principle. Regular
sampling of position allows us to determine momentum below its
Nyquist limit, and similarly the regular sampling of momentum
will allow us to determine position below its Nyquist limit.
%
%

%
\end{document}